\documentclass[conference]{IEEEtran}
\IEEEoverridecommandlockouts

\usepackage{cite}
\usepackage{amsmath,amssymb,amsfonts}
\usepackage{algorithmic}
\usepackage{graphicx}
\usepackage{textcomp}
\usepackage{xcolor}
\def\BibTeX{{\rm B\kern-.05em{\sc i\kern-.025em b}\kern-.08em
    T\kern-.1667em\lower.7ex\hbox{E}\kern-.125emX}}

\begin{document}

\title{Graph Neural Network based Service Function Chaining for Automatic Network Control\\
\thanks{This work was supported by Institute for Information and communications Technology Promotion (IITP) grant funded by the Korea government (MSIT) (No.2018-0-00749, Development of virtual network management technology based on artificial intelligence), and partly by the Basic Science Research Program through the National Research Foundation of Korea (NRF) funded by the Ministry of Education (2017R1D1A1B03033341).}} 

\author{\IEEEauthorblockN{DongNyeong Heo}
\IEEEauthorblockA{
\textit{Dept. of Information and Communication Engineering} \\
\textit{Handong Global University}\\
Pohang, South Korea \\
21931011@handong.edu}
\and
\IEEEauthorblockN{Stanislav Lange}
\IEEEauthorblockA{
\textit{Dept. of Information Security and Communication Technology} \\
\textit{Norwegian University of Science and Technology}\\
Trondheim, Norway \\
stanislav.lange@ntnu.no}
\and
\IEEEauthorblockN{Hee-Gon Kim}
\IEEEauthorblockA{
\textit{Dept. of Computer Science and Engineering} \\
\textit{Pohang University of Science and Technology}\\
Pohang, South Korea \\
sinjint@postech.ac.kr}
\and
\IEEEauthorblockN{Heeyoul Choi}
\IEEEauthorblockA{\textit{Dept. of Information and
Communication Engineering} \\
\textit{Handong Global University}\\
Pohang, South Korea \\
heeyoul@gmail.com}
}

\maketitle

\begin{abstract}
Software-defined networking (SDN) and the network function virtualization (NFV) led to great developments in software based control technology by decreasing expenditures. Service function chaining (SFC) is an important technology to find efficient paths in network servers to process all of the requested virtualized network functions (VNF). However, SFC is challenging since it has to maintain high Quality of Service (QoS) even for complicated situations. Although some works have been conducted for such tasks with high-level intelligent models like deep neural networks (DNNs), those approaches are not efficient in utilizing the topology information of networks and cannot be applied to networks with dynamically changing topology since their models assume that the topology is fixed. In this paper, we propose a new neural network architecture for SFC, which is based on graph neural network (GNN) considering the graph-structured properties of network topology. The proposed SFC model consists of an encoder and a decoder, where the encoder finds the representation of the network topology, and then the decoder estimates probabilities of neighborhood nodes and their probabilities to process a VNF. In the experiments, our proposed architecture outperformed previous performances of DNN based baseline model. Moreover, the GNN based model can be applied to a new network topology without re-designing and re-training.

\end{abstract}

\begin{IEEEkeywords}
Service Function Chaining, Deep Learning, Graph Neural Network
\end{IEEEkeywords}

\section{Introduction}
Reducing Capital Expenditure (CAPEX) and Operating Expenditure (OPEX) are consistently critical issues for telecommunication network service providers (NSPs). Before appearances of Software-Defined Network (SDN) and Network Function Virtualization (NFV), network functions were dependent on hardware middleboxes. The SDN technology \cite{kreutz:sdn-survey} led to network traffic being controlled by a software-based control system. In addition, the NFV technology \cite{mijumbi:nfv-survey} led to network functions being virtualized and make them be separated from hardware middleboxes. With SDN and NFV technologies, NSPs are capable of deploying and processing Virtualized Network Functions (VNF) at a relatively low cost in terms of CAPEX and OPEX. These developments have led the software-based control system to receive much attention from NSPs. Also, high-level intelligent software-based control systems that operate networks automatically have become a central issue in the network field \cite{datta:intelligent-networking}.

\begin{figure}
    \centering
    \includegraphics[width=0.83\linewidth]{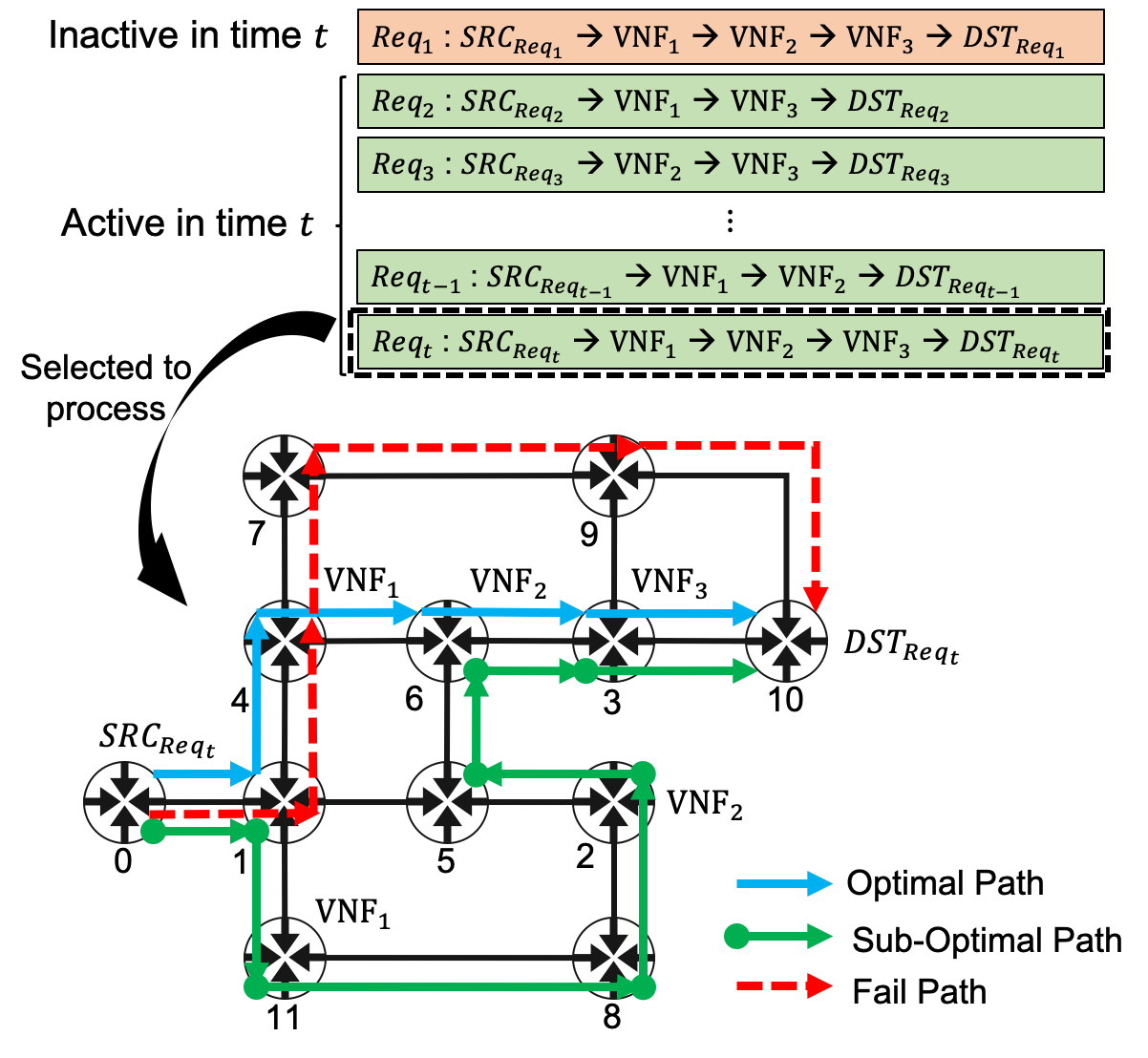}
    \caption{An example of an SFC task on network topology. The request list is updated by new or expired requests. The SFC model generates resulting paths for the active requests in the list. The bottom figure shows the generated SFC paths of one request.}
    \label{fig:sfc}
\end{figure}

A high-level intelligent software-based control system needs to handle many specific tasks including VNF Deployment, VNF Placement, Auto-Scaling, and Anomaly Detection. Among them, the Service Function Chaining (SFC) task generates a network traffic path from the source server to the destination server. At the same time, connecting and processing all of the VNFs that are requested from users in a pre-defined order. To guarantee Quality of Service (QoS) for the SFC, there are a few requirements such as low end-to-end delay, enough bandwidth to operation, reliability, and availability. In this study, the QoS is evaluated by total delay costs of traversing traffic between nodes. Even with a single QoS measurement, SFC is still challenging since the network should satisfy the good quality of QoS even in complicated situations like dynamic locations of VNF instances, various request types from different users, and a structure-level change of the network topology.

To formulate the task, we can consider network topology as graph data. Then, physical servers are represented by nodes in the graph, and paths to traverse network traffic between physical servers are represented by edges. Fig. \ref{fig:sfc} shows a simple example of network topology and an SFC task. The request list is updated whenever new requests are received or existing requests expire. In the figure, a new request $Req_t$ is received at time $t$ and the first request $Req_1$ has expired. The SFC model generates a path for each of the active requests in the list. To make the task simpler, requests are processed one by one, not simultaneously. Suppose that the active request $Req_t$ is selected for processing as shown in Fig. \ref{fig:sfc}, then given the pre-defined order of requested VNFs, the model generates a path from the source node (node 0) to the destination node (node 10) processing all the requested VNFs (VNF$_1$, VNF$_2$, and VNF$_3$). Depending on the QoS of the resulting path, it could be an optimal or sub-optimal, or even a failure path. The resulting path is classified as a failure when the path does not process all the requested VNFs.

Threshold methods for SFC are limited regarding their capability of handling complicated network situations \cite{Bhamare:sfc-survey}. Recently, deep learning-based models have been applied to SFC, since deep learning has received much attention over the last decade because of its success in various domains like image recognition, speech recognition, and machine translation \cite{goodfellow:deeplearningbook}. Also, some researches have attempted to apply deep learning to SFC. \cite{pei:deep-sfc} designed a pre-trained Deep Belief Network (DBN) model that estimates the probabilities of neighborhood nodes after fine-tuning. 
Iterative steps of forwarding DBN model generate a path. This research shows that the deep learning-based model can learn a representation of the network topology. However, its representation does not reflect the relationships of the nodes in the network topology, since those relationships cannot be reflected by DBN.

In this paper, we propose a new neural network architecture for SFC based on graph neural networks (GNNs) \cite{scarselli:gnn} followed by deep neural networks (DNNs) or recurrent neural networks (RNNs). GNN is advantageous for graph-structured data because of its effective representation of the relationship between graph nodes. Moreover, GNN can be applied without re-designing even when the structure of network topology is changed. We designed an SFC model using the encoder-decoder architecture, where the encoder represents the network topology, and the decoder estimates the probabilities of neighborhood nodes for traversing. At the same time, the decoder estimates the probabilities of processing the deployed VNF. In the experiments, our proposed GNN based architecture outperforms the DNN based model which is used as a baseline. Furthermore, experiment results show that the GNN based model can be applied to a new network topology without re-designing.

\section{Background}

\subsection{Service Function Chaining}
In this paper, for the definition of the SFC task, we follow \cite{pei:deep-sfc} which is also used as a baseline model in our experiments. 
Also, we take several notations from \cite{scarselli:gnn, li:ggnn}, and they have different meanings with the same notations. Thus, for consistency of notations in our task with GNN equations, we use our own notations. First of all, the graph $G$ consists of a node set $N$ and an edge set $E$. An instance of the node set and the edge set are denoted by $u$ and $uv$, respectively, where $uv$ represents the connection between the two nodes $u$ and $v$. The set of VNFs is denoted by $M$ whose instance is denoted by $m$.
Basically, the network topology in SFC is considered to be an undirected graph, but each step of the resulting path can be understood as a directed graph.
In the directed graph, a node and an edge are denoted by $\bar{u}$ and $\bar{u}\bar{v}$, respectively. Lastly, the active request list at time $t$ is denoted by $R_t$. 

The objective of SFC for a request $i \in R_t$ is defined by
\begin{equation}
    \min \sum_{uv \in E}\sum_{\bar{u}\bar{v} \in E_i} d_{uv}y_{i,uv}^{\bar{u} \bar{v}} + \sum_{m \in M}\sum_{\bar{u} \in N_i} d_m x_{i,m}^{\bar{u}}, \nonumber
\end{equation}
where $d_{uv}$ is the traversing delay on the edge $uv$ and $d_m$ is the processing delay on the VNF instance $m$. Given a generated path of the request $i$, $N_i$ and $E_i$ are the set of nodes and the set of directed edges of the generated path. $\bar{u}$ and $\bar{u}\bar{v}$ are instances of such sets. $y_{i,uv}^{\bar{u}\bar{v}}$ is a variable indicating whether the directed edge $\bar{u}\bar{v}$ traverses the undirected edge $uv$. $x_{i,m}^{\bar{u}}$ indicates whether the node $\bar{u}$ processes the VNF instance $m$ in the path. The first term of this equation is the sum of traversing delays on edges, and the other one is the sum of processing delays on the VNF deployed nodes. 
Also, each VNF instance and link in which between nodes have their own bandwidth capacities which should be considered when generating the path. See \cite{pei:deep-sfc} for more details of notations and constraints.

\subsection{Graph Neural Networks (GNNs)}
GNNs were proposed to handle graph-structured data \cite{scarselli:gnn} in neural networks. The traditional purpose of the graph processing model is to learn a function that maps a node $u$ in a graph $G$ into a vector representation of real numbers, $\tau(G,u) \in \mathbb{R}^D$ where $D$ is the dimension of the vector. Before GNNs, most neural networks were not effective in utilizing graph topology information. Vector representation about relations between a node and its neighborhood is necessary to handle graph-structured data in neural networks. GNNs have a state transition stage producing a state representation that reflects the information of relations. Additionally, GNNs reflect features that are referred to as label information of nodes or edges in the state transition stage. In the SFC task, we set the label of a node and edge as the type of deployed VNFs and inverse delay cost of that edge.

\begin{figure}[t]
    \hbox{
        \hspace{0.5in}\includegraphics[scale=0.33]{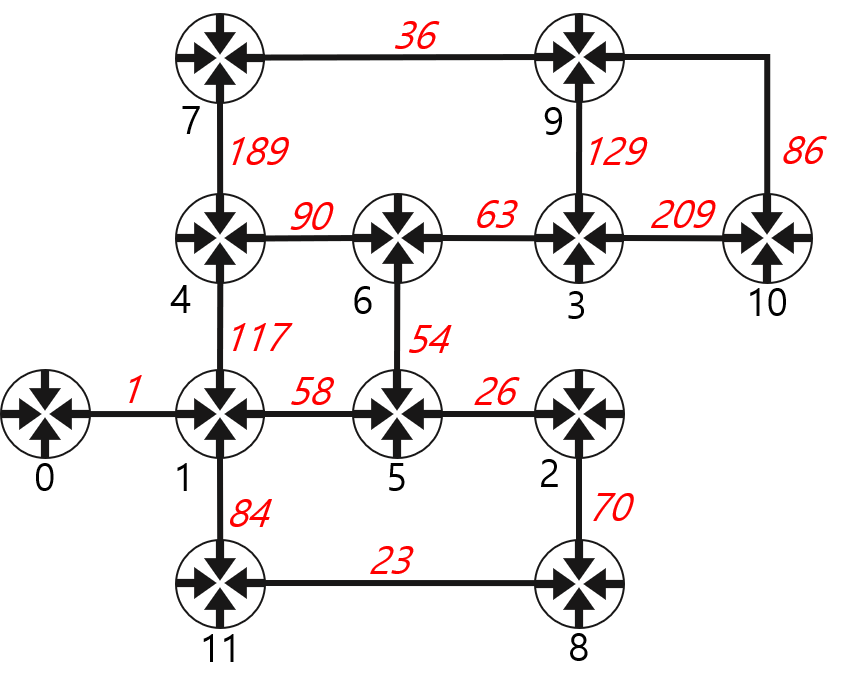}
    }
    \hbox{\small \hspace{0.55in}(a) Internet2 Topology and Settings}
    \vspace{0.2in}
    \hbox{\hspace{0.5in}
        \includegraphics[scale=0.26]{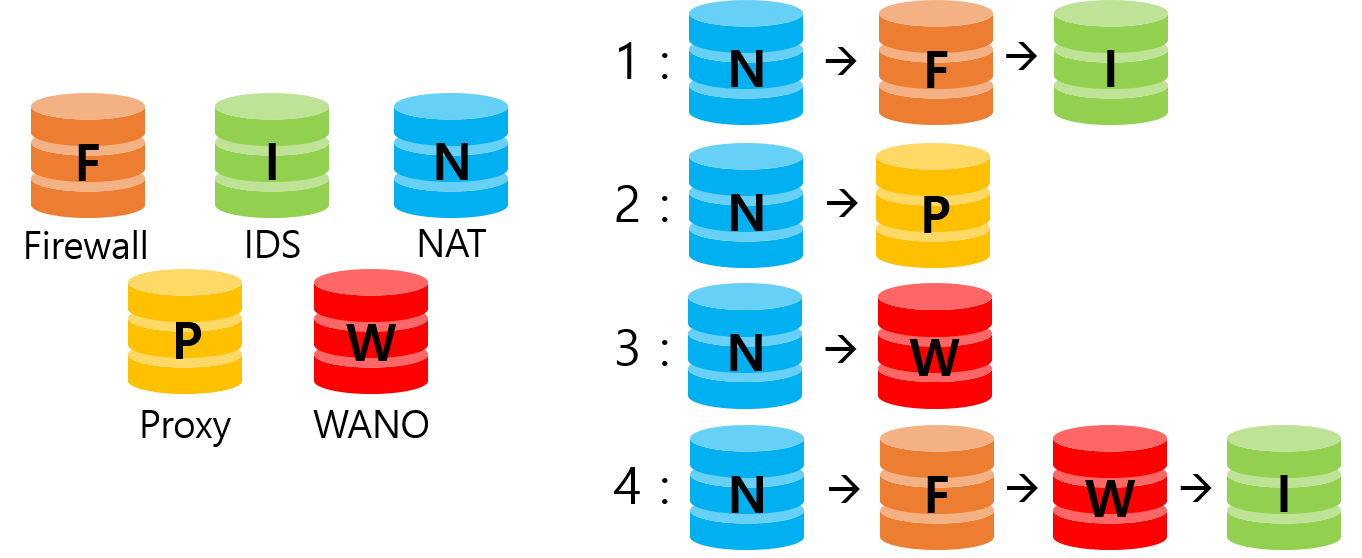} 
    }
    \hbox{\small \hspace{0.3in}(b) VNF Types (Left) and SFC Request Types (Right)}
\caption{Data collecting network topology, types of VNF, and user request.}
\label{fig:internet2}
\end{figure}

Let $ne[u]$ be the neighbors of node $u$, and $co[u]$ the edge set that connects the node $u$ to any $ne[u]$. Nodes and edges have their own label, expressed in the vector format like $l_u \in \mathbb{R}^{l_N}$, $l_{uv} \in \mathbb{R}^{l_E}$, where $l_N$ and $l_E$ are dimensions of the node and edge labels, respectively. For an SFC task, we assume that the graph topology is an undirected and non-positional graph.
The GNN model is divided into two main stages. The first stage is the state transition as mentioned before, which produces a state representation reflecting the information of relations. This stage is expressed as follows.
\begin{equation}
    h_u = \sum_{v \in ne[u]}  f_w(l_u,l_{uv},h_v,l_v),
    \label{eq:f_local}
\end{equation}
where $f_w$ with a parameter set $w$ is a transition function and $h_u$ is the state representation of the node $u$. It reflects the individual information of relations computed between $u$ and one of its neighbors $ne[u]$. The GNN model repeats this state transition Eq. \eqref{eq:f_local}, until the output of the function converges. The Banach fixed-point theorem guarantees the function to find the unique solution independently to the initial parameter set under the condition that the function is a contraction map. To satisfy this condition, a regularizer can be adopted as in \cite{mijumbi:gnn-deploy}. 

The second stage is output function which produces the final output vector representation of a node given the state representation from the first stage, and this output is given by 
\begin{equation}
    o_u = g_w(h_u,l_u),
    \label{eq:f_output}
\end{equation}
where $o_u$ is the final output vector representation of the node $u$, and $g_w$ is the output function with a parameter set $w$.

\subsection{Gated Graph Neural Network (GG-NN)}
Variants of GNN have been proposed with different neural network architectures, such as Convolutional Neural Networks \cite{kipf:gcn} and Recurrent Neural Networks (RNNs) \cite{li:ggnn}. Especially in \cite{li:ggnn}, they implemented GNN with RNNs based on the idea that the forwarding of the state transition stage is the same process as the forwarding of RNN models with masking between a hidden state and the next hidden state. GG-NN applied Gated Recurrent Unit (GRU) \cite{cho:gru} to GNN.

In GG-NN, the state transition stage is implemented as the matrix multiplication of the hidden state matrix and the adjacency matrix. The initial hidden state matrix is the annotation matrix that is a set of node label vectors, and the adjacency matrix is a set of edge label vectors. The operations of this state transition stage are summarized as follows.
\begin{align}
    h_u^{(0)}&=[l_u^\top,0]^\top, \label{eq:init_h} \\
    a_u^{(t)}&=A_u^\top[h_1^{(t-1)\top} \dots h_{|N|}^{(t-1)\top}]^\top, \label{qe:init_adj} \\
    z_u^t&=\sigma(W^za_u^{(t)}+U^z h_u^{(t-1)}), \label{eq:gru_z} \\
    r_u^t&=\sigma(W^ra_u^{(t)}+U^rh_u^{(t-1)}), \label{eq:gru_r} \\
    \widetilde{h}_u^{(t)}&=\tanh{(Wa_u^{(t)}+U(r_u^t \odot h_u^{(t-1)}))}, \label{eq:gru_htilde} \\
    h_u^{(t)}&=(1-z_u^t) \odot h_u^{(t-1)} + z_u^t \odot \widetilde{h}_u^{(t)}, \label{eq:gru_newh}
\end{align}
where $A_u$ is the vector of the node $u$ in the adjacency matrix. $h_u^{(0)}$ is the vector of the node $u$ in the annotation matrix with zero paddings. $a_u^{(t)}$ is multiplication of the adjacency matrix and annotation matrix. $z_u^t, r_u^t, \widetilde{h}_u^{(t)}$ and $h_u^{(t)}$ are usual recursive state transition operations in GRU. $\odot$ is the element-wise multiplication operation.
Restricting the model to be a contraction map might lose the expressive power of the model. Therefore, GG-NN repeats the recursive state transition stage with fixed times \cite{li:ggnn} without the restriction of contraction. In this paper, we adopt the GG-NN to build our neural network architecture for the SFC task.

\begin{figure*}[t]
    \hbox{\hspace{-0.0in}
        \includegraphics[scale=0.33]{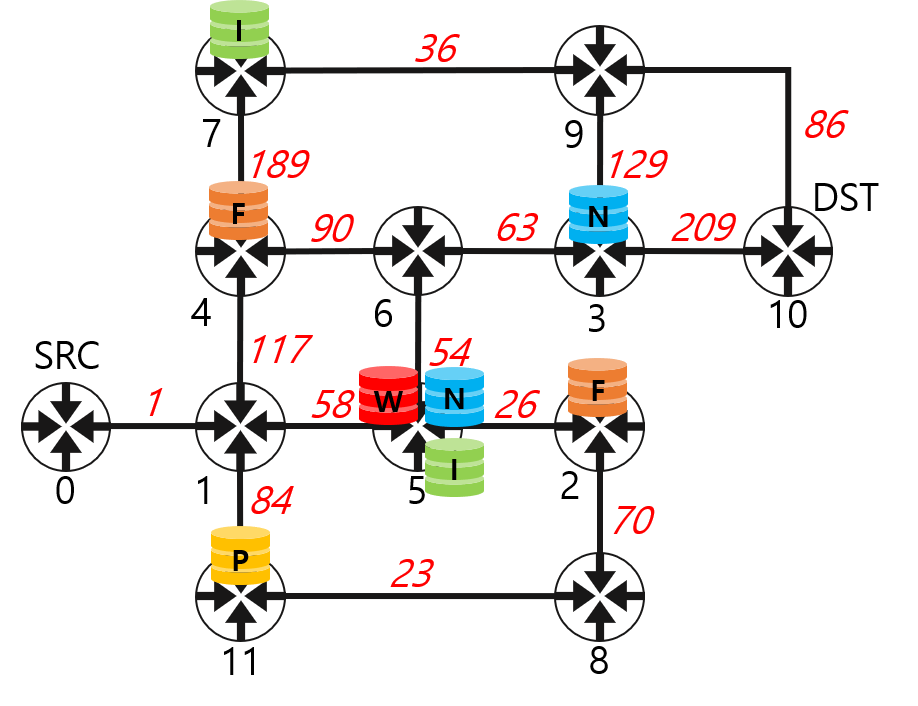}
    \hspace{0.1in}
        \includegraphics[scale=0.39]{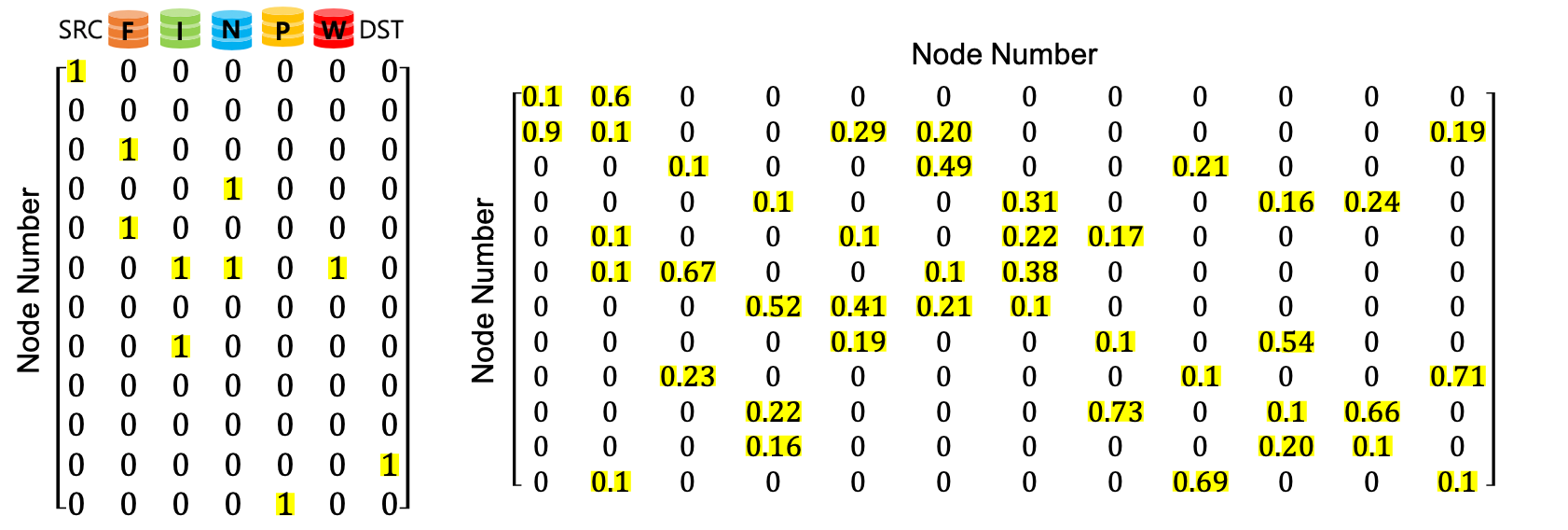}
        }
    \hbox{\small \hspace{0.4in}(a) Network topology setting \hspace{1.4in}(b) Annotation matrix (Left), and adjacency matrix (Right)}
    \caption{An example of the annotation matrix and adjacency matrix given the network topology with 12 nodes and 5 VNF types.}
    \label{fig:annotation&adjacency}
\end{figure*}

\section{GNN-Based Service Function Chaining}
The network topology we used in the experiment is Internet2 as illustrated in Fig. \ref{fig:internet2}(a). There are traversing delays for each edge as presented in red tilted. Fig. \ref{fig:internet2}(b) shows 5 VNF types and specific orders of SFC request types. With this network topology, we propose a GNN-based SFC model, which consists of two submodels: an encoder and a decoder. Detailed descriptions of these submodels are below.

\subsection{Encoder}
The annotation matrix is a set of label vectors with the features of a node, including a special type of that node, such as a source node, destination node, or VNF deployed node. For example, Fig. \ref{fig:annotation&adjacency}(a) is one of the network topology setting of Fig. \ref{fig:internet2}(a), and the annotation matrix is illustrated in Fig. \ref{fig:annotation&adjacency}(b) (Left). The matrix size ($12 \times 7$) is determined by 12 nodes and 7 features that include 5 types of VNF, source, and destination.
If a node has a specific VNF type, then the dimension of the vector is set to one, otherwise zero. For example, the node 5 has `I', `N' and `W' types, so the 6th row has $0 0 1 1 0 1 0$. Since the annotation matrix elements are symbolic, we need embedding process. The embedding process is similar to word embedding in language models \cite{bengio:language_model}, which finds a distributed representation vector that may represent many independent factors \cite{mikolov:word2vec}.

The adjacency matrix is a set of label vectors indicating features of edges as shown in Fig. \ref{fig:annotation&adjacency}(b) (Right). The matrix elements can be obtained by the inverse of traversing delay cost of the edges, which is $A_{uv}=1/d_{uv}$. Then, each column is normalized with its mean and standard deviation values, then the softmax function is applied to each column so that the total amount of information emitting from a node can be one. If an edge is not connected, the corresponding element become zero after softmax.  

\begin{figure}[t]
    \hbox{\centering \hspace{0.15in}
    \includegraphics[width=0.7\linewidth]{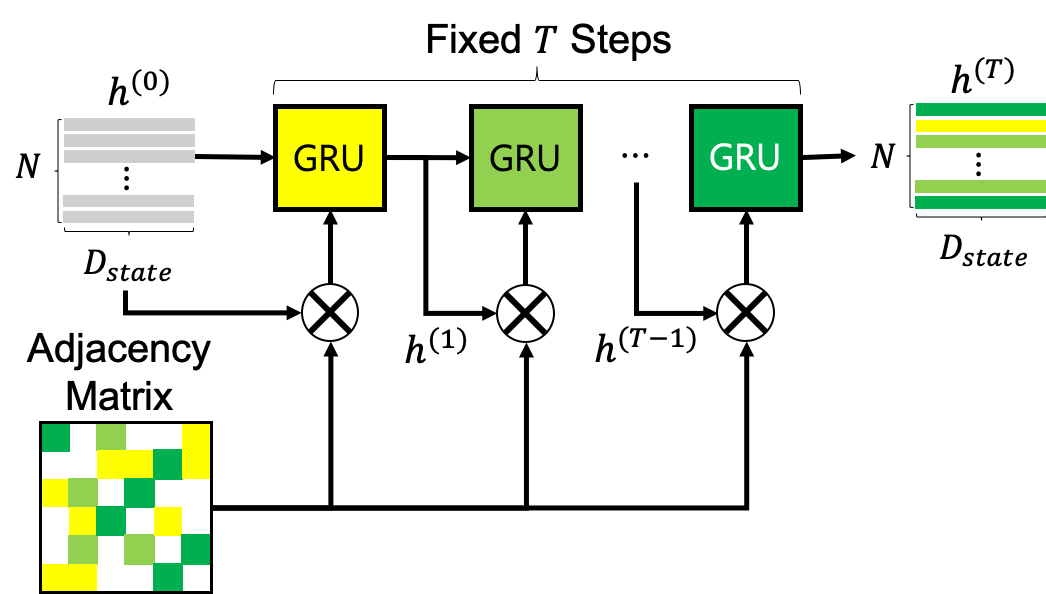}
    }
    \hbox{\small \hspace{1.0in}(a) Encoder Model}
    \vspace{0.1in}
    \hbox{\centering \hspace{-0.075in}
    \includegraphics[width=0.98\linewidth]{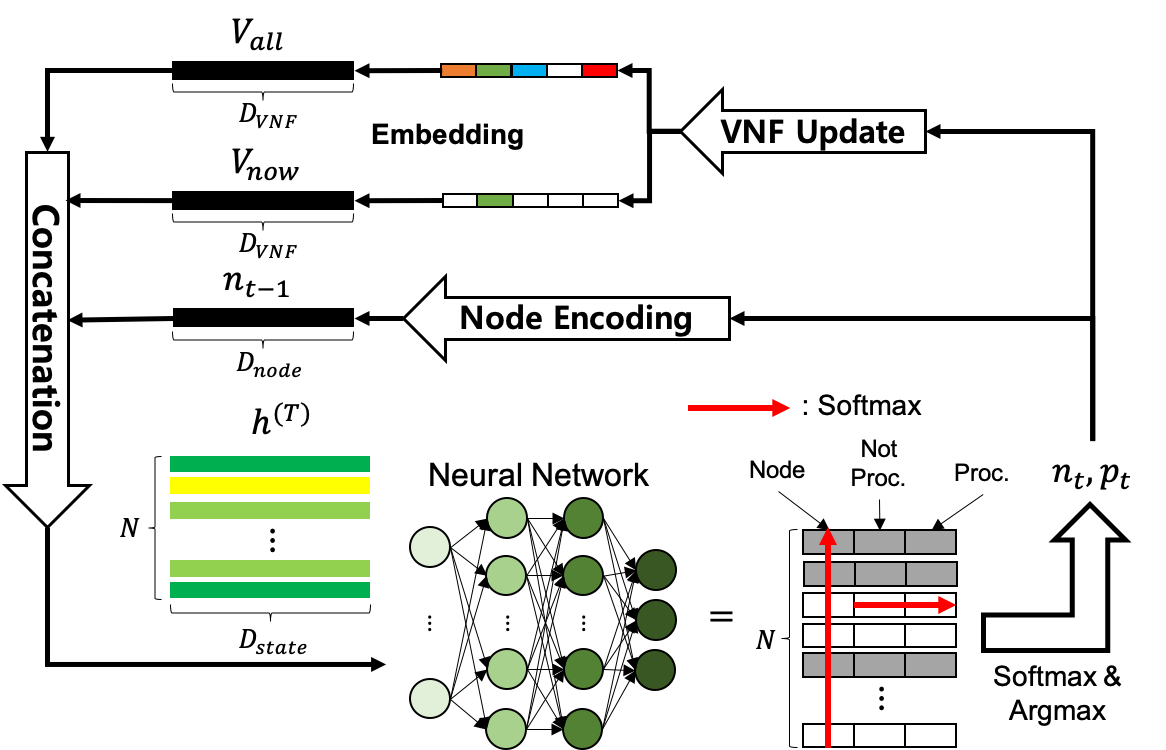}
    }
    \hbox{\small \hspace{1.0in}(b) Decoder Model}
    \caption{Encoder and Decoder models for SFC. In the encoder, strength of adjacency matrix elements is presented in dark color.}
    \label{fig:model_architecture}
\end{figure}

The encoder plays the same role as the GNN state transition stage as shown in Eq. \eqref{eq:f_local}. The encoder produces a state representation that reflects the information of relations between nodes. Based on the GG-NN architecture, the encoder can be summarized as
\begin{equation}
    h_u^{(t)}=f_w^{enc}(a_u,h_u^{(t-1)}), 
    \label{eq:f_w_enc}
\end{equation}
where $f_w^{enc}$ is a summary of the GG-NN processes from Eq. \eqref{eq:gru_z} to Eq. \eqref{eq:gru_newh}. The encoder repeats these processes for fixed $T$ times. Fig. \ref{fig:model_architecture}(a) illustrates the entire encoder model. After recursive state transitions for $T$ times, the information of relations is presented in the final state representation $h^{(T)}$.

\subsection{Decoder}
To generate a path for a request considering the final representations of the encoder, the decoder finds one node at a time until it completes a path. To complete the path, it decides whether to process the VNF on the selected node or not. This decoding process is similar to the language model or neural machine translation processes \cite{bengio:language_model}. At each decoding step, to select the next node and VNF process, the decoder estimates the probabilities of the neighbor nodes and their probabilities to process the VNF.

As shown in Fig. \ref{fig:model_architecture}(b) since the selection needs to reflect the current context of the path in the network topology, three additional inputs from the network topology are given.  Eq. \eqref{eq:f_w_enc}: (1) the whole list of requested VNF types $V_{all}$, (2) the next VNF type that the SFC model should process currently $V_{now}$, and (3) the current node that the model is located currently $n_{t-1}$. The VNF type vectors $V_{all}$ and $V_{now}$ have five dimensions as the number of VNF types, and then they are processed with embedding. The location vector $n_{t-1}$ is computed by positional encoding (PE) via sine and cosine functions with different frequencies. PE is the same technique as in Transformer \cite{vaswani:transformer} where PE is used to identify location information of words in neural machine translation. By applying VNF vector embedding and node encoding, the three additional input vectors ($V_{all}$, $V_{now}$, and $n_{t-1}$) are transformed and have new dimensions $D_{VNF}$, $D_{VNF}$, and $D_{node}$, respectively.

The decoder estimates two types of probabilities for the neighbor nodes: one for selecting the next node and the other for processing the VNF on the node. The output vector $o_u$ has three dimensions. For all $u \in ne[n_{t-1}]$, the probability $p(u)$ and $p(proc|u)$ is obtained by
\begin{align}
    o_u&=f_w^{dec}(h_u^{(T)},V_{all},V_{now},n_{t-1}),\label{eq:decoder_output} \\
    p(u)&=\dfrac{e^{o_u^{node}}}{\sum_{v\in{ne[n_{t-1}]}}e^{o_v^{node}}}, \label{eq:decoder_prob}, \\
    p(proc|u)&=\dfrac{e^{o_{u}^{proc}}}{e^{o_{u}^{proc}}+e^{o_{u}^{not}}}. \label{eq:decoder_prob_process}
\end{align}
Note that the probabilities are computed only for the neighbor nodes of the previous node, $ne[n_{t-1}]$. 

Fig. \ref{fig:model_architecture}(b) describes the decoder model architecture. The additional inputs that are followed by embedding and encoding are concatenated with the encoded representation matrix from the encoder. This concatenated input is fed to the decoding neural network $f_w^{dec}$, and the neural network estimates the probabilities of the neighbor nodes and their probabilities to whether processing the VNF or not. Then, masked softmax and argmax operator select the target node $n_t$ for the next step. The scores of processing VNF on the selected node $n_t$ is computed with softmax function for binary classification. Then, for the next step, the generated node $n_t$ and the decision of processing VNF update the additional input vectors $V_{all}$, $V_{now}$ and $n_{t-1}$. In training, instead of the generated node $n_t$, the true labels of both node and processing VNF are used as the next step input which is teacher forcing manner \cite{bengio:teacher_forcing}.

\subsection{Integrated Model}
Our proposed model integrates the encoder-decoder architecture and it is end-to-end trainable by backpropagation. The objective function consists of two cross-entropy (CE) terms: one for the target node classification, and the other for the binary classification indicating whether processing the VNF on the selected node or not. The encoder is implemented as GG-NN and the encoded representation forms a vector set whose size is the number of nodes $|N|$ $\times$ the dimension of the hidden states. The trained model can be applied to new network topologies without re-designing or re-training even when the number of nodes changes. The encoder based on GG-NN can handle changing the number of nodes in the topology, and the decoder treats the changing number of nodes as if it is changing batch size. If the encoder is implemented as a DNN rather than GNN, it should be re-designed when the number of input dimensions is changed. This is an important advantage of our proposed model about to dynamic network topologies.

\section{Experiments and Results}
\subsection{Data Description}

Following the data format in \cite{lange2019predicting}, we created a dataset from the Internet2 topology and dynamic SFC requests. For each request, we used an ILP-based placement algorithm to determine the optimal number and location of VNF instances to generate paths for the set of active requests of that time. We used this ILP-based result as the label to train the proposed neural network model. We have a total of 14735 lists for the active requests, and each list contains 26.08 requests on average. We split the dataset into 13135, 100, and 1500 for training, validation, and testing, respectively. See \cite{lange2019predicting} for more details about data generations and structures.

\subsection{Models and Hyperparameters}
We designed a baseline model based on multi-layered DNN which is a modified version of the pre-trained DBN model \cite{pei:deep-sfc}. We changed the hyperbolic tangent activation function to the ReLU function. In addition, we added several optimization techniques such as dropout, learning rate decay, and early stopping \cite{hahn2020}. Also, we increased the model complexity of the baseline model for a fair comparison with our proposed model. The total number of layers is 4 and each layer has 256 hidden states. This baseline model receives every piece of information about the network topology, the annotation status of each node, and their relationships (traversing delay costs). Also, the three additional information which is the same as additional inputs of our proposed decoder model is given to the baseline model. Then, the baseline model estimates same output probabilities with our proposed model in every generating step.

The encoder in the proposed model is based on GG-NN and the encoding step $T$ is fixed to 5. The hidden state size $D_{state}$ and the embedded annotation vector dimension are set to 128. The decoder in the proposed model has a neural network as illustrated in Fig. \ref{fig:model_architecture}(b), which is based on either DNN or GRU, leading to GG-DNN or GG-RNN respectively. For the DNN based decoding model, the neural network has 4 hidden layers with the ReLU activation function and dropout regularizer. For the GRU based decoding, the neural network has 256 hidden state dimensions. For the additional information, embedded VNF input vectors $D_{VNF}$, and positional encoding vector $D_{node}$ are represented in 32 and 4 dimensions, respectively. The maximum length of the path is limited to 50, so if the decoder generates a longer path, then it is considered as failure. We optimized the model with RMSprop optimizer with an initial learning rate of 0.0001. The total number of parameters is 626K, 579K, and 530K for the DNN baseline, GG-DNN, and GG-RNN, respectively.

\subsection{Evaluation metric}
To evaluate the performance for validation and testing, we could not use the accuracy metric, since the lengths of the target path and the generated path are different. Instead, we calculated the average cost ratio between the generated path $\hat{p}$ and the true path $p$, $\frac{1}{|P|} \sum_{p \in P} \frac{cost(\hat{p})}{cost(p)}$, where $P$ is the set of true paths, and $cost(p)$ is the total delay cost of the path. We evaluated the mean and variance of the ratio.

Also, we checked the number of failures over the total path generations, `Fail Ratio'. Specific cases that the model cannot generate a path to process all the requested VNFs, when the remaining resource was insufficient, or when the resource was sufficient but the model could not find an available path are considered as failures. 
Lastly, we also checked the rate of unsuccessful cases `Overmax' in which the total delay cost of the generated path was higher than the pre-defined maximum delay for the request.

\begin{table}[htbp]
\caption{Test results with 3 models for SFC}
\begin{center}
\begin{tabular}{|l||c|c|c|}
\hline
\textbf{Model}& \textbf{\textit{Avg. Cost Ratio (Var.)}} &\textbf{\textit{Fail Ratio}} &\textbf{\textit{Overmax}}\\
\hline \hline
DNN (baseline) & 1.209 (0.729) & 0.063 & 0.243   \\
\hline
GG-DNN  & 1.008  (0.542) & 0.038 & 0.152      \\
\hline
GG-RNN  & 0.995  (0.504) & 0.012 & 0.159     \\
\hline
\end{tabular}
\label{tab:test_results}
\end{center}
\end{table}

\subsection{Results}
Table \ref{tab:test_results} summarizes the results showing that both GG-DNN and GG-RNN outperform the baseline in terms of the mean of the cost ratio. Also, the `Fail Ratio' of GG-RNN is 1.2\% which is significantly better than the baseline and GG-DNN. In addition, our proposed model performed better (0.159) then the ILP-based solution (0.236, not shown in the table) for the `Overmax' evaluation metric. This means that our model was able to find more effective solutions than the target label in certain cases. Although further investigation is needed to understand the reason why the proposed model showed a lower ratio for the `Overmax' than the ILP-based solution. We believe it is related to the fact that our model generates paths for requests in a list one by one while the target labels were found all at once. Because the one by one method consumed resources of VNF instances without consideration of later requests, the proposed model could generate shorter paths while increasing the `Fail Ratio'.

\begin{table}[htbp]
\caption{Test Results on Changed Topology}
\begin{center}
\begin{tabular}{|l||c|c|c|}
\hline
\textbf{Model}& \textbf{\textit{Avg. Cost Ratio (Var.)}} &\textbf{\textit{Fail Ratio}} &\textbf{\textit{Overmax}}\\
\hline \hline
DNN (baseline) & \multicolumn{3}{|c|}{\textbf{Not applicable}}    \\
\hline
GG-DNN  & 1.148 (0.698) & 0.664 & 0.069      \\
\hline
GG-RNN  & 1.078 (0.535) & 0.128 & 0.188     \\
\hline
\end{tabular}
\label{tab:test_results_on_changed_topology}
\end{center}
\end{table}

To prove that the proposed models can work even when the internet topology changes, we tested the models after changing the topology setting by adding nodes that connect distant nodes with low delay costs. Two nodes were added newly for connecting nodes 0 and 9, also nodes 8 and 10 respectively. As shown in Table \ref{tab:test_results_on_changed_topology}, the results of GG-RNN on the changed topology show that the model can be applied to the changed topology without retraining while the DNN model could not be applied. Although the proposed model for the changed topology was not able to as effective as the original topology, the GG-RNN model could find paths for 87\% of whole requests and the generated paths almost as good as the true paths (mean of the ratio is 1.078). We believe that the lower performance (especially in Fail Ratio) on the changed topology is due to overfitting, that is, the training samples were drawn only from a single topology and the model was optimized excessively for that topology. If the model is trained on various topologies, then we expect that the model works effectively on new topologies without overfitting.

\section{Conclusion}
To find an SFC path on internet networks automatically, we proposed new neural network architectures based on the graph neural network via encoder-decoder architecture. The encoder found representations of the network topology, and the decoder estimated probabilities of the neighbor nodes and the probabilities of decisions of processing VNF, then the decoder makes the best choices given that probabilities. In the experiments, the proposed models had not only a higher level of performance than the baseline model but also demonstrated flexibility in structural changes of topology without re-designing. For future work, we can train the model on various topologies to avoid the overfitting issue.

\bibliographystyle{IEEEtran}
\bibliography{my_bib}

\end{document}